\documentstyle[12pt,epsf]{article}
\def \S{S$\chi$PT }
\def \G{G$\chi$PT }
\def \sub{subthreshold coefficients }
\newcommand{\Frac}[2]{\frac{\displaystyle #1}{\displaystyle #2}}

\begin{document}
\pagestyle{empty}
\begin{titlepage}
\begin{center}
\vspace*{-2cm}
\hfill DTP-95/84 \\
\hfill INFNNA-IV-96/39 \\
\hfill DSFNA-IV-96/39 \\
\hfill September, 1996
\vspace*{0.8cm} \\
{\LARGE \bf Standard vs. Generalized \\
Chiral Perturbation Theory~: 
\vspace*{0.4cm} \\
the Chell-Olsson Test}
\vspace*{1.0cm}\\
{\large \bf M.R. Pennington} 
\vspace*{0.1cm} \\
Centre for Particle Theory, \\
University of Durham \\
Durham DH1 3LE, U.K. 
\vspace*{0.3cm} \\
and
\vspace*{0.3cm} \\
{\large \bf J. Portol\'es} 
\vspace*{0.1cm} \\
INFN, Sezione di Napoli \\
Dipartimento di Scienze Fisiche, Universit\`a di Napoli \\
I-80125 Napoli, Italy 
\vspace*{0.5cm} \\
\begin{abstract}
A way of testing the $\pi \pi$ predictions of Chiral Perturbation Theory
against experimental data is to use dispersion relations to continue
experimental information into the subthreshold region where the theory
should unambiguously apply. Chell and Olsson have proposed a test of 
the subthreshold behaviour of chiral expansions which highlights potential
differences between the Standard and the Generalized forms of the theory.
We illustrate how, with current experimental uncertainties, data cannot
distinguish between these particular {\em discriminatory} coefficients
despite their sensitivity. Nevertheless, the Chell--Olsson test does
provide a consistency check of the chiral expansion, requiring that
the ${\cal O}(p^6)$ corrections to the {\em discriminatory} coefficients
in the Standard theory must be $\sim 100\%$. Indeed, some of these
have been deduced from the new ${\cal O}(p^6)$ computations and found
to give such large corrections. One can then check that the 
${\cal O}(p^8)$ corrections must be much smaller. 
\par
We conclude that this test, like others, cannot distinguish between
the different forms of Chiral Symmetry Breaking embodied in the alternative
versions of Chiral Perturbation Theory without much more precise experimental
information near threshold.
\end{abstract}
\end{center}
\end{titlepage}
\newpage
\pagestyle{plain}
\pagenumbering{arabic}
\baselineskip=7.5mm

\section{Introduction}
\hspace*{0.5cm}

\baselineskip=7.5mm
\parskip=2.5mm

The fact that scattering amplitudes
are analytic functions means that their behaviour
 at different energy scales are related. 
Chiral dynamics controls low energy pion reactions and, 
for instance, requires 
that the amplitude for $\pi^+\pi^-\to\pi^0\pi^0$ has a line of real zeros below
threshold.  This on-shell manifestation of the Adler zero 
within the Mandelstam triangle, in turn demands  that the 
$\pi^+\pi^0\to\pi^+\pi^0$ amplitude
must grow asymptotically. Such relationships between  the behaviour of
scattering amplitudes at different energies are naturally embodied in dispersion
relations.  These can be used as a way of expressing subthreshold amplitudes 
as integrals over
physical region absorptive parts, to be determined either experimentally 
or theoretically \cite{paver}.
Chiral Perturbation Theory ($\chi$PT)  allows the same subthreshold
quantities to be expressed directly in terms of the parameters of the 
Chiral Lagrangian.
There are two realizations of $\chi$PT~: Standard (S$\chi$PT) \cite{gasleut}
and Generalized
(G$\chi$PT) \cite{stern}. In S$\chi$PT there are two expansion parameters~: 
the momentum 
squared of an emitted pion and the pion mass characterizing the explicit
breaking of chiral symmetry.  In G$\chi$PT the quark condensate
matrix element is regarded as an additional dimensionful parameter, in terms 
of which the standard chiral
expansion is reordered.  At any finite order either of S$\chi$PT and G$\chi$PT
may have an expansion with smaller higher order corrections.
\par
The predictions of $\chi$PT can be compared with the evaluation of dispersion
relations in two different ways, which depend on the inputs to the
dispersive integrals. In an idealized Test~A, the absorptive parts are 
input wholly from experiment, then the comparison of the subthreshold
expansion coefficients with the predictions of $\chi$PT tests the efficacy
of the chiral expansion to some given order. Alternatively, in Test~B the
absorptive parts are input from $\chi$PT (at least at low energies). Then
the comparison tests that the amplitudes of $\chi$PT satisfy the
appropriate analyticity and crossing properties, fulfil unitarity at 
least perturbatively and are consistent with experiment for energies
beyond where $\chi$PT applies. We will consider these two inequivalent
tests in turn.
\par
At the 1994 workshop on {\it Chiral Dynamics} at MIT, Olsson \cite{mit}
 presented the first of these as
a \lq\lq stringent test" of the chiral expansion schemes, initially 
reported in the thesis of Chell.
Guided by experimental data, Chell and Olsson evaluated the subthreshold
 expansion coefficients (to be formally defined in Sect. 2) using
dispersion relations and compared these with the predictions of $\chi$PT 
in both
its standard and generalized forms.  While many coefficients 
evaluated from experiment agreed with both
versions of $\chi$PT, several evaluated from experiment  were found to be in 
far better agreement
with G$\chi$PT (with smaller quark condensate) typically by a factor of 2.
 The results of this test are so
intriguing that this issue is worth investigating further.
\par
A number of questions immediately come to mind~:
\begin{itemize}
{\leftskip 1.5cm\rightskip 1.5cm{
\item[(i)] does the better agreement with G$\chi$PT depend on the choice of
experimental input~?
\item[(ii)] what is particular about the coefficients that are the basis of
this discriminatory test~?

}}
\end{itemize}
These questions, among others, are what we answer in this paper. 
In Sect.~2, we define the subthreshold expansion and the dispersive 
representation
of the corresponding coefficients. In Sect.~3 we give the explicit 
evaluation of these coefficients at ${\cal O}(p^4)$ $\chi$PT in its two 
forms. In Sect.~4 we compute dispersively these same coefficients using
a flexible parametrization of low energy $\pi \pi$ scattering. We then
compare the dispersive and explicit evaluations of the subthreshold
coefficients, which allows us to discuss the accuracy of the 
${\cal O}(p^4)$ chiral expansions. We shall see, however, that this Test~A
is inconclusive because of the sizeable experimental uncertainties in 
the near threshold amplitudes. In Sect.~5 we turn to Test~B, which 
checks the consistency of the chiral expansions at any given order.
In Sect.~6 we present our conclusions.

\section{Defining the Tests}
\hspace*{0.5cm}
The predictions of $\chi$PT can be verified in two ways. Either the 
predictions can be continued into the physical regions, where data 
exist, but then one is uncertain about what energy regime is really
appropriate for a given order in $\chi$PT, or, by using dispersion
relations, experimental data can be continued below threshold, where
$\chi$PT should unambiguously apply. The latter is what we do here
by considering the $\pi \pi$ amplitude in the Mandelstam triangle.
\par
To this end, we consider the amplitudes with definite isospin in 
the $t$-channel~: $A^{I_t}(s,t,u)$. From these we construct the
functions $\widetilde{F}^{I_t}(\nu,t)$, where
\begin{equation}
\widetilde{F}^{I_t}(\nu, t)\;=\; A^{I_t}(s,t,u) 
\quad\qquad{\rm for}\quad I_t=0,2
\label{eq:fit1}
\end{equation}
\noindent and
\begin{equation}
\widetilde{F}^{I_t}(\nu, t)\;=\; A^{I_t}(s,t,u)/\nu 
\quad\quad {\rm for} \quad I_t=1\,
\label{eq:fit2}
\end{equation}
\noindent with
\begin{equation}
 \nu\,=\,{{s-u}\over{4\mu^2}}\qquad {\rm and}\qquad s + t + u = 4\mu^2\; 
\label{eq:nu}
\end{equation}
\noindent and $\mu=m_{\pi}$, the pion mass \footnote{Note that this definition
of $\nu$ differs from that of Chell and Olsson, who use $\nu = (s-u)/4\mu$.}.
The three amplitudes $\widetilde{F}^{I_t}$ are  symmetric under 
$\nu \to -\nu$. 
\par
Now, rather than work with these amplitudes throughout the Mandelstam triangle,
it is more convenient to study their Taylor series
expansion about the subthreshold point $t=0$, $\nu=0$ (i.e. $s=u=2 \mu^2$)~:
\begin{equation}
\widetilde{F}^{I_t}(\nu, t)\;=\;\sum_{k,m}\,
F^{(I_t)}_{k,m}\,\nu^{2k}\,t^m\qquad ,
\label{eq:texpa}
\end{equation}
and to study the coefficients $F^{(I_t)}_{k,m}$.
\par
Regge theory leads us to expect that for $\mid \nu \mid \to \infty$
at fixed $t$~:
\vspace{1.5mm}
\begin{eqnarray}
\widetilde{F}^{I_t=0} (\nu, t) & \;\sim\; & \nu^{\alpha_P(t)}
\qquad{\rm where}\qquad 
\alpha_P(0) \simeq 1.08\quad\; , \nonumber  \\
\label{eq:regge}
\widetilde{F}^{I_t=1} (\nu, t) & \;\sim\; & \nu^{\alpha_{\rho}(t)-1}\quad\,
{\rm where}\qquad 
\alpha_{\rho}(0) \simeq 0.5 \quad , \\
\widetilde{F}^{I_t=2} (\nu, t) & \;\sim\; & \nu^{\alpha_E(t)}
\qquad{\rm where}\qquad 
\alpha_E(0) < 0\qquad . \nonumber
\end{eqnarray}

\noindent Consequently, $\widetilde{F}^{I_t=1,2}$ satisfy unsubtracted 
dispersion relations,
while that for $\widetilde{F}^{I_t=0}$ requires one subtraction for 
$t \leq 4\mu^2$. Writing $F^{I_t}(s,t) \equiv \widetilde{F}^{I_t}(\nu,t)$
we have
\begin{equation}
F^{I_t}(s,t)\;=\;{1\over\pi}\,\int_{4\mu^2}^{\infty}
\;ds'\,\left( {1\over{s'-s}}\,+\,{1\over{s'-u}}\, \right)\,
{\rm Im}\ F^{I_t}(s',t) \; \; \;  ,
\label{eq:f12dr}
\end{equation}
\noindent for $I_t = 1,2$, while
\begin{eqnarray}
F^{I_t=0}(s,t)\;& = & \;F^{I_t=0}(2\mu^2-t/2,t)  
\label{eq:f0dr} \\
& & \,+\,{1\over\pi}\,
\int_{4\mu^2}^{\infty}
\;ds'\,\left( {1\over{s'-s}}\,+\,{1\over{s'-u}}
-\, {2\over{s'-2\mu^2+t/2}}\right)\,
{\rm Im}\ F^{I_t=0}(s',t)\, . \nonumber
\end{eqnarray}
\noindent We are primarily interested in the subthreshold coefficients,
$F^{(I_t)}_{k,m}$, for which the dispersive integral is dominated by the 
low energy
absorptive parts.  For $I_t = 1,2$, this means $k\,+\,m\,\ge 1$, while for
$I_t =0$ to avoid the dependence on the subtraction term in Eq. 
(\ref{eq:f0dr}) also requires
at least one derivative with respect to $\nu^2$, i.e. $ k \ge 1$.  We 
therefore consider

\begin{eqnarray}
F^{(I_t)}_{1,0} \,&  = & \, \Frac{8 \mu^4}{\pi} \, \int_{4 \mu^2}^{\infty} \,
\Frac{ds'}{(s'-2 \mu^2)^3} \; {\rm Im} F^{I_t} (s',0) \; \; \; \; ,
\label{eq:fit10} \\
& & \nonumber \\
\mu^2 \, F^{(I_t)}_{1,1} \, &  = & \, \Frac{4 \mu^6}{\pi} \, 
\int_{4 \mu^2}^{\infty} \, \Frac{ds'}{(s' - 2 \mu^2)^3} \, 
\Bigg( \, 2 \, \Frac{\partial}{\partial t} \, {\rm Im} F^{I_t} (s',t) 
\, \Big|_{t=0} \,
- \, \Frac{3}{s'-2 \mu^2} \, {\rm Im} F^{I_t} (s',0) \Bigg)  ,
\label{eq:fit11} \\
& & \nonumber \\
F^{(I_t)}_{2,0} \, & = & \, \Frac{32 \mu^8}{\pi} \, \int_{4 \mu^2}^{\infty} \, 
\Frac{ds'}{(s' - 2 \mu^2)^5} \; {\rm Im} F^{I_t} (s',0)
\label{eq:fit20}
\end{eqnarray}

\noindent for $I_t=0,1,2$, and

\begin{eqnarray}
\mu^2 F^{(I_t)}_{0,1} \, & =  & \, \Frac{\mu^2}{\pi} \, 
\int_{4 \mu^2}^{\infty} \, \Frac{ds'}{s' - 2 \mu^2} \, \Bigg(
\, 2 \, \Frac{\partial}{\partial t} \, {\rm Im} F^{I_t} (s',t) \, \Big|_{t=0} 
\, - 
\, \Frac{1}{s' - 2 \mu^2} \, {\rm Im} F^{I_t} (s',0) \, \Bigg)   \; ,
\label{eq:fit01} \\
& & \nonumber \\
\mu^4 F^{(I_t)}_{0,2} \, &  = & \, \Frac{\mu^4}{\pi} \, \int_{4 \mu^2}^{\infty} 
\, \Frac{ds'}{s'- 2 \mu^2} \, \Bigg( \, \Frac{\partial^2}{\partial t^2} \, 
{\rm Im} F^{I_t} (s',t) \, \Big|_{t=0} \, - \, \Frac{1}{s' - 2 \mu^2} \, 
\Frac{\partial}{\partial t} {\rm Im} F^{I_t} (s',t) \, \Big|_{t=0} \, 
\nonumber \\
 &  & \qquad \qquad \qquad \qquad 
+ \Frac{1}{2 (s' - 2 \mu^2)^2} \, {\rm Im} F^{I_t} (s',0) \Bigg)   
\label{eq:fit02}
\end{eqnarray}

\noindent for $I_t = 1,2$.
\par
These form the basis of the Chell--Olsson tests in  the forms 
previously mentioned. Either we input the experimental data for the
$\pi \pi$ amplitudes in the dispersive integrals to determine the 
\sub (Test~A), 
or we input the $\chi$PT amplitudes to do so (Test~B) \footnote{ Of course,
Test~B  can only be carried out if the bulk of the contribution to the 
dispersive integrals comes from the very low energy region where 
$\chi$PT safely applies. As we will see, this is the case
for the so called {\it discriminatory} coefficients.}.
\newpage
\section{Explicit evaluation of subthreshold coefficients}
\hspace*{0.5cm}
 We next compute the subthreshold coefficients, defined by 
Eq.~(\ref{eq:texpa}), in Standard
 and Generalized $\chi$PT based on the formulae of Refs. 
\cite{gasleut,stern} for the ${\cal O}(p^4)$ $\pi \pi$ amplitudes. We obtain 
\vspace*{0.5cm} \\
\noindent {\underline{\bf S$\chi$PT}}

 \noindent ${\underline {I_t=0}}$~: 
\begin{eqnarray}
 F^{(0)}_{1,0}\,& = & \,{1\over{18432\pi^3}}\,\left({\mu\over{F_{\pi}}}
\right)^4\,
\left[ 476 -183\pi +96 (\overline{\ell}_1 + 4 \overline{\ell}_2 ) \right] 
\; \;  , 
\label{eq:f010} \\
& & \nonumber \\
\mu^2 F^{(0)}_{1,1} \, &  = & \, \Frac{1}{8192 \pi^3}\, \left(
\Frac{\mu}{F_{\pi}} \right)^4\, \left[ 88 - 61 \pi \right] \; \; , 
\label{eq:f011} \\
& & \nonumber \\
F^{(0)}_{2,0} \, & = & \, \Frac{1}{24576 \pi^3} \, \left(
\Frac{\mu}{F_{\pi}} \right)^4 \, \left[ 608 - 135 \pi \right] \; \; , 
\label{eq:f020}
\end{eqnarray}

\noindent ${\underline {I_t=1}}$~:
\begin{eqnarray}
F^{(1)}_{1,0} \, &  = & \, \Frac{7}{18432 \pi^3} \, \left(
\Frac{\mu}{F_{\pi}} \right)^4 \, \left[ 8 + 3 \pi \right] \; \; , 
\label{eq:f110} \\
& & \nonumber \\
\mu^2 F^{(1)}_{0,1} \, & = & \, \Frac{1}{36864 \pi^3} \, \left(
\Frac{\mu}{F_{\pi}} \right)^4 \, \left[ 76 - 87 \pi + 
96 ( \overline{\ell}_2 - \overline{\ell}_1 ) \right] \; \; , 
\label{eq:f101} \\
& & \nonumber \\
\mu^2 F^{(1)}_{1,1} \, &  = &  \, \Frac{1}{73728 \pi^3} \, \left(
\Frac{\mu}{F_{\pi}} \right)^4 \, \left[ 64 - 93 \pi \right] \; \; , 
\label{eq:f111} \\
& & \nonumber \\
F^{(1)}_{2,0} \, &  = & \, \Frac{1}{368640 \pi^3} \, \left(
\Frac{\mu}{F_{\pi}} \right)^4 \, \left[ 512 + 75 \pi \right] \; \; , 
\label{eq:f120} \\
& & \nonumber \\
\mu^4 F^{(1)}_{0,2} \, & = & \, \Frac{1}{163840 \pi^3} \, \left(
\Frac{\mu}{F_{\pi}} \right)^4 \, \left[ 328 - 45 \pi \right] \; \; , 
\label{eq:f102}
\end{eqnarray}

\noindent ${\underline {I_t=2}}$~:
\begin{eqnarray}
 F^{(2)}_{1,0}\, & = & \,{1\over{18432\pi^3}}\,\left({\mu\over{F_{\pi}}}
\right)^4\,
\left[ 51 \pi - 76 + 96 (\overline{\ell}_1 +  \overline{\ell}_2 ) \right] 
\; \;  , 
\label{eq:f210} \\
& & \nonumber \\
\mu^2 F^{(2)}_{0,1} \, &  = & \, \Frac{1}{32 \pi} \, \left( 
\Frac{\mu}{F_{\pi}} \right)^2 \, \left\{ -1 \, + \, 
\Frac{1}{1152 \pi^2} \, \left( \Frac{\mu}{F_{\pi}} \right)^2  
\left[ 130 + 39 \pi - 192 \overline{\ell}_2 - 144 \overline{\ell}_4
\right] \, \right\}  \; , 
\label{eq:f201} \\
& & \nonumber \\
\mu^2 F^{(2)}_{1,1} \, & = & \, \Frac{-1}{8192 \pi^3} \, \left(
\Frac{\mu}{F_{\pi}} \right)^4 \, \left[ 8 + 19 \pi \right] \; \; , 
\label{eq:f211} \\
& & \nonumber \\
F^{(2)}_{2,0} \, & = & \, \Frac{1}{24576 \pi^3} \, \left(
\Frac{\mu}{F_{\pi}} \right)^4 \, \left[ 32 + 27 \pi \right] \; \; , 
\label{eq:f220} \\
& & \nonumber \\
\mu^4 F^{(2)}_{0,2} \, & = & \, \Frac{1}{491520 \pi^3} \, \left(
\Frac{\mu}{F_{\pi}} \right)^4 \, \left[ 160 ( \overline{\ell}_1 +
5 \overline{\ell}_2 ) - 468 - 75 \pi \right] \; \; .
\label{eq:f202} 
\end{eqnarray}

\noindent In Eqs.~(\ref{eq:f010}-\ref{eq:f202}) $\overline{\ell}_1, \, 
\overline{\ell}_2$ and $\overline{\ell}_4$ are effective coupling 
constants that appear in the polynomial part of the ${\cal O} (p^4)$ 
Chiral Lagrangian in S$\chi$PT \cite{gasleut}.
\vspace*{0.2cm} \\
\noindent {\underline{\bf G$\chi$PT}} 
\\
\noindent ${\underline {I_t=0}}$~:

\begin{eqnarray}
F^{(0)}_{1,0} \, & = & \, \Frac{\mu^4}{9} \, \left( 120 \alpha_0 +
16 \beta_0 \right) \, + \, \Frac{3 \beta^2}{\pi (96 \pi F_{\pi}^2)^2} \left\{
\, \left( 8 - 2 \pi \right) \kappa_0^2 \, + \, \left(
10 - \Frac{5}{2} \pi \right) \kappa_2^2 \right. \nonumber \\
& & \; \qquad \qquad \qquad \qquad \qquad \qquad \qquad \; \; \; 
 + \,  8 \pi \mu^2 \kappa_0 \, + \, 10 \pi \mu^2 \kappa_2
 \label{eq:fg010} \\
& & \qquad \qquad \qquad \qquad \qquad \qquad \qquad \; \; \; \left.  
\;  +  \, \left( 112 + 4 \pi \right) \mu^4  \right\} \;  , 
 \nonumber \\
& & \nonumber \\
\mu^2 F^{(0)}_{1,1} \, & = & \, \Frac{3}{8 \pi} \, 
\Frac{\beta^2}{(96 \pi F_{\pi}^2)^2} \, \left\{ \, ( 32 - 12 \pi ) 
\kappa_0^2 \, + \, (40 - 15 \pi) \kappa_2^2 \, - \, 64 \mu^2 \kappa_0 
\, \, \right. \nonumber \\
& & \qquad \qquad \qquad \; \; \left.  - \, 80 \mu^2 \kappa_2 \, 
- \, \left( 120 \pi + 96 \right) 
\mu^4 \, \right\} \; \; , \label{eq:fg011} \\
& & \nonumber \\
F^{(0)}_{2,0} \, & = & \, \Frac{1}{8 \pi} \, 
\Frac{\beta^2}{(96 \pi F_{\pi}^2)^2} \, \left\{ \, (128 - 36 \pi) 
\kappa_0^2 \, + \, ( 160 - 45 \pi ) \kappa_2^2 \, + \, 48 \pi \mu^2
\kappa_0 \, \right. \nonumber \\
& & \qquad \qquad \qquad \; \; \left. + \, 60 \pi \mu^2 \kappa_2 \, + \, 
(1152 - 72 \pi) \mu^4 \, 
\right\} \; \; , \label{eq:fg020}
\end{eqnarray}

\noindent ${\underline {I_t=1}}$~:

\begin{eqnarray}
F^{(1)}_{1,0} \, & = & \, \Frac{1}{4 \pi} \, 
\Frac{\beta^2}{(96 \pi F_{\pi}^2)^2} \, \left\{ \, (24 \pi - 64) 
\kappa_0^2 \, + \, (40 - 15 \pi) \kappa_2^2 \, + \, 128 \mu^2 \kappa_0\,
\, \, \right. \nonumber \\
& & \qquad \qquad \qquad \; \; \left.  - \, 80 \mu^2 \kappa_2 \, 
+ \, ( 96 \pi - 128 ) \mu^4 \, \right\}  \; \; , 
\label{eq:fg110} \\
& & \nonumber \\
\mu^2 F^{(1)}_{0,1} \, & = & \, \Frac{4}{3} \mu^4 \beta_0 \, 
  + \, \Frac{\beta^2}{\pi (96 \pi F_{\pi}^2)^2} \left\{
\, \left( 3 \pi - 12 \right) \kappa_0^2 \, + \, \Frac{15}{8} \left(
4 - \pi  \right) \kappa_2^2 \right. \, \nonumber \\
& & \qquad   \qquad \qquad \qquad \qquad \left. - 
\, 12 \pi \mu^2 \kappa_0
 +  \Frac{15}{2} \pi \mu^2 \kappa_2  -  \left( 20 + 6 \pi \right) 
\mu^4  \right\} \;  , 
\label{eq:fg101} \\
& & \nonumber \\
\mu^2 F^{(1)}_{1,1} \, & = & \, \Frac{1}{16 \pi}
 \Frac{\beta^2}{(96 \pi F_{\pi}^2)^2} \left\{ \,
( 9 \pi - 32 ) (8 \kappa_0^2 - 5 \kappa_2^2 ) \, + \, 
12 \pi \mu^2 ( 5 \kappa_2 - 8 \kappa_0 ) \right. \nonumber \\
& & \qquad \qquad \qquad \; \; \left. + \, 16 \mu^4 ( 3 \pi - 28 ) \right\} \;
\; , \label{eq:fg111} \\
& & \nonumber \\
F^{(1)}_{2,0} \, & = & \, \Frac{1}{80 \pi} 
 \Frac{\beta^2}{(96 \pi F_{\pi}^2)^2} \left\{ \,
(45 \pi - 128 ) (8 \kappa_0^2 - 5 \kappa_2^2 ) \, + \, 
128 \mu^2 ( 8 \kappa_0 - 5 \kappa_2 ) \right. \nonumber \\
& & \qquad \qquad \qquad \; \; \left. + \, 64 \mu^4 ( 15 \pi - 32 )
\right\} \; \; , \label{eq:fg120} \\
& & \nonumber \\
F^{(1)}_{0,2} \, & = & \, \Frac{3}{320 \pi} 
 \Frac{\beta^2}{(96 \pi F_{\pi}^2)^2} \left\{ \,
5 (3 \pi - 8 ) ( 8 \kappa_0^2 - 5 \kappa_2^2 ) \, + \, 
80 \mu^2 (8 \kappa_0- 5 \kappa_2 ) \, \right. \label{eq:fg102} \\
& & \qquad \qquad \qquad \; \; \; \; \left.  + \, 768 \mu^4 \, 
\right\} \; \;  ,
\nonumber
\end{eqnarray}

\noindent ${\underline {I_t=2}}$~:

\begin{eqnarray}
F^{(2)}_{1,0} \, & = & \, \Frac{8}{9} \mu^4 ( 6 \alpha_0 - \beta_0 ) 
\, + \,  \Frac{3 \beta^2}{4 \pi (96 \pi F_{\pi}^2)^2} \left\{ 
 \left( 4 - \pi  \right) ( 8 \kappa_0^2 + \kappa_2^2 )
\, + \, 4 \pi \mu^2 ( 8 \kappa_0 + \kappa_2 ) \right. \nonumber \\
& & \qquad \qquad \qquad \qquad \qquad \qquad \; \; \;  \left. 
\; \; \; \; \; \; + \, 64 \left( 1 + \pi  \right) \mu^4 \right\} \; \; ,
\label{eq:fg210} \\
& & \nonumber \\
\mu^2 F^{(2)}_{0,1} \, & = & \, - \Frac{\mu^2}{3} \beta_2 \, - \, 
\Frac{8}{9} \mu^4 \left( 3 \alpha_0 + \beta_0 \right) \, - \, 
\Frac{\beta \mu^2}{32 \pi F_{\pi}^2} \nonumber \\
& & \, + \, \Frac{3 \beta^2}{\pi ( 96 \pi F_{\pi}^2 )^2} \, \left\{
\,  \left( 2 - \pi \right) \kappa_0^2 \, + \, 
\Frac{1}{8} \left( 6 - \pi  \right) \kappa_2^2 \, - \, 
 \mu^2 ( \kappa_2 + 8 \kappa_0 )  \right. \label{eq:fg201} \\
& & \qquad \qquad \qquad \; \; \left. + \, 4 \left( \pi - 6 
\right) \mu^4 \right\} \nonumber \\
& & \nonumber \\
\mu^2 F^{(2)}_{1,1} \, & = & \, \Frac{3}{16 \pi} 
 \Frac{\beta^2}{(96 \pi F_{\pi}^2)^2} \left\{ \,
(8 - 3 \pi) (8 \kappa_0^2 + \kappa_2^2 ) \, - \, 16 \mu^2 ( 8 \kappa_0 
+ \kappa_2 ) \right. \nonumber \\
& & \qquad \qquad \qquad \; \left. \; - \, 96 ( 2 + \pi ) \mu^4 \right\}
\; \; , \label{eq:fg211} \\
& & \nonumber \\
F^{(2)}_{2,0} \, & = & \, \Frac{1}{16 \pi}
 \Frac{\beta^2}{(96 \pi F_{\pi}^2)^2} \left\{ \,
(32 - 9 \pi ) ( 8 \kappa_0^2 + \kappa_2^2 ) \, + \, 
12 \pi \mu^2 ( 8 \kappa_0 + \kappa_2 ) \right. \nonumber \\
& & \qquad \qquad \qquad \; \; \left. + \, 288 \pi \mu^4 \right\} 
\; \; , \label{eq:fg220} \\
& & \nonumber \\
\mu^4 F^{(2)}_{0,2} \, & = & \, \Frac{\mu^4}{6} ( 6 \alpha_0 
+ \beta_0 ) \, + \, \Frac{3}{320 \pi} 
 \Frac{\beta^2}{(96 \pi F_{\pi}^2)^2} \left\{ \,
40 ( 4 - \pi ) \kappa_0^2 \, + \,  ( 36 - 5 \pi ) \kappa_2^2 \right. 
\nonumber \\
& & \qquad \qquad \qquad \qquad \qquad \qquad \qquad 
\, + \, 160 \pi \mu^2 \kappa_0 \, + \, 20 ( 16 + \pi ) \mu^2 \kappa_2 
\, \label{eq:fg202} \\
& & \qquad \qquad \qquad \qquad \qquad \qquad \qquad \left.
\, + \, 1280 \mu^4 \right\} \, . \nonumber
\end{eqnarray}

\noindent In Eqs.~(\ref{eq:fg010}-\ref{eq:fg202}) $\alpha_0, \, \beta_0$ and
$\beta_2$ are parameters that depend on $\alpha, \, \beta$ and subtraction
constants in the dispersive analysis of Stern et al. \cite{stern}, while
\begin{eqnarray}
\kappa_0 \, \doteq \, \left( \Frac{5 \alpha}{6 \beta} - \Frac{4}{3} \right) 
\mu^2 & , & \; \kappa_2 \, \doteq \, -  \left( 
\Frac{2\alpha}{3\beta} + \Frac{4}{3} \right) \mu^2 \; \; . 
\label{eq:kappas}
\end{eqnarray} 
 
As an aside, we note that the subthreshold coefficients, Eq.~(\ref{eq:texpa}),
 are not in fact
 independent. While their definition embodies the $s-u$ symmetry of the 
amplitudes
 (as in $\pi K$ or $\pi N$ scattering), the $\pi\pi$ process actually 
has three-channel crossing. 
 This means that the three isospin amplitudes can each be written in terms 
of one
 function, e.g. the Chew-Mandelstam invariant amplitude $A(s,t,u)$. This
 imposes conditions among the $F^{(I_t)}_{k,m}$. For instance,
\begin{eqnarray}
F^{(2)}_{0,0}\,& = &\, \;\, {1\over 3} \left( F^{(0)}_{0,0} - F^{(2)}_{0,0} 
\right)
 \,+\,
{4\over 3} \left( F^{(0)}_{0,1} - F^{(2)}_{0,1} \right) \nonumber \\
& &\,+\,{8\over 3} \left( F^{(0)}_{0,2} - F^{(2)}_{0,2} \right)\,+\,
{1\over 6} \left( F^{(0)}_{1,0} - F^{(2)}_{1,0} \right) \nonumber \\
& &\,+\,{1\over 24} \left( F^{(0)}_{2,0} - F^{(2)}_{2,0} \right)
\,+\,\sum_{k,m>2}\,c_{k,m}\, \left( F^{(0)}_{k,m} - F^{(2)}_{k,m} \right)  
\label{eq:tricros1} \\
& & \nonumber \\
F^{(2)}_{0,1}\,& = &\,-\, {1\over 3} \left( F^{(0)}_{0,1} - F^{(2)}_{0,1} 
\right) \,-\,
{4\over 3} \left( F^{(0)}_{0,2} - F^{(2)}_{0,2} \right) \nonumber \\
& & \;-\,{1\over 4} \left( F^{(0)}_{1,0} - F^{(2)}_{1,0} \right)\,-\,
{1\over 8} \left( F^{(0)}_{2,0} - F^{(2)}_{2,0} \right) \nonumber \\
& & \;+\,\sum_{k,m>2}\,d_{k,m}\, \left( F^{(0)}_{k,m} - F^{(2)}_{k,m} \right) 
\label{eq:tricros2}
\end{eqnarray} 
 \noindent In S$\chi$PT $\mu^2 F^{(2)}_{0,1}\,=\,-2.6 \times  10^{-2}$, 
while the
 $k + m \leq 2$ terms in Eq.~(\ref{eq:tricros2}) give $-2.7 \times 10^{-2}$.
  In contrast for
 $F^{(2)}_{0,0}$, Eq.~(\ref{eq:tricros1}), S$\chi$PT gives 
$5.0 \times 10^{-2}$, whereas the $k + m \leq 2$ terms give 
 $6.6 \times  10^{-2}$. So these relationships from three-channel 
crossing are not in practice very useful, since they require connections between
a large number of coefficients : relationships that are, of course, 
automatically
satisfied by any crossing symmetric representation, like that of $\chi$PT.
\par 
Of the coefficients listed in Eqs.~(\ref{eq:f010}-\ref{eq:f202}), we see 
in S$\chi$PT that apart from
 $F^{(0,2)}_{1,0}$, $ F^{(1)}_{0,1}$, $F^{(2)}_{0,1}$ and $F^{(2)}_{0,2}$, 
the others do not depend on the
 $\overline{\ell_i}$, which specify the  polynomial (resonance
generated) ${\cal O}(p^4)$  corrections to 
the Chiral
 Lagrangian. Contrastingly \cite{gasleut} the $I=0$ $S$--wave scattering
length
\begin{equation}
a_0^0\,=\,{{7\mu^2}\over{32\pi F_{\pi}^2}}\,\left\{
 1\,+\,{5\over{84\pi^2}} \left(\mu\over F_{\pi} \right)^2\, \left[
 \overline{\ell}_1 \,+\, 2 \overline{\ell}_2 \, - 
 \,{3\over 8} \overline{\ell}_3 \, + \, \Frac{21}{10} \overline{\ell}_4 \,
 +\,{21\over 8}\,\right]\,\right\}
\label{eq:a00chpt}
\end{equation}
on which, as we shall see the dispersive integrals crucially depend, does
involve the $\overline{\ell}_i$.
In G$\chi$PT at ${\cal O}(p^4)$, the coefficients all depend on
$\alpha, \beta$, as does 
$a^0_0$, Eqs.~(\ref{eq:fg010}--\ref{eq:kappas}) 
\cite{pipigchpt}, in the following way~:
\vspace{1mm}
\begin{equation}
a_0^0 \, = \, \Frac{\mu^2}{96 \pi F_{\pi}^2}\, \left\{
\,(5 \alpha + 16 \beta)\,\left(\, 1 + \Frac{\mu^2}{48 \pi^2 F_{\pi}^2}
(5 \alpha + 16 \beta)
\, \right) \, + \, 60 \left( \Frac{\mu}{F_{\pi}} \right)^2 ( \lambda_1
+ 2 \lambda_2 ) \, \right\} \, \; ,
\label{eq:a00gchpt}
\end{equation}
\vspace*{0.1cm} \\
\noindent  where $\lambda_1$ and $\lambda_2$ can be written in terms
of the $\overline{\ell}_i$'s of S$\chi$PT as
\vspace{2mm}
\begin{eqnarray}
\lambda_1 \, = \, \Frac{1}{48 \pi^2} \, \left( \overline{\ell}_1 - 
\Frac{4}{3} \right) \; \; & , & \; \; 
\lambda_2 \, = \, \Frac{1}{48 \pi^2} \, \left( \overline{\ell}_2 - 
\Frac{5}{6} \right) \; \; . 
\label{eq:lambdas}
\end{eqnarray}
 \\
 We now evaluate the \sub, $F^{(I_t)}_{k,m}$, using the following 
set of parameters~\footnote{These are the values of the parameters 
for S$\chi$PT and 
G$\chi$PT quoted by Martin Olsson in his talk at the MIT workshop 
\cite{mit} .}~:
\vspace{2mm}
\begin{eqnarray}
\overline{\ell}_1\,\,=\,\,-1.1\;\qquad & , & \qquad \overline{\ell}_2\,
=\,5.7\quad ,\nonumber \\
 \overline{\ell}_3\,\,=\, \, \; \,\,2.9\;\qquad & , & \qquad \overline{\ell}_4\,
=\,1.6\quad , \label{eq:lis} 
\end{eqnarray}
 
\noindent for S$\chi$PT.
As is well--known, if $\alpha=\beta=1$ the ${\cal O}(p^2)$ \G is identical
to its Standard form. This remains approximately true at higher orders if
$\alpha \approx 1$, $\beta \approx 1$. In \G, while $\beta$ is always
close to $1$, $\alpha$ is roughly between 
$1$ and $4$ depending on the
magnitude of the quark condensate. Since we want to compare and in particular
contrast the two versions of $\chi$PT, we here take \G to have~$^3$ :
\vspace{2mm}
\begin{eqnarray}
\alpha\;=\; 3.1\;\; \; \; \; \qquad & , &\qquad\; \;\;\;\;
 \;\beta\,=\,0.93\quad , \nonumber \\
 \alpha_0\ \mu^4\,=\,5.5 \times  10^{-4}\, \,  & , &\qquad \beta_0\ \mu^4\,
=\,3.5 \times 10^{-3}\; ,   \label{eq:gchptpar} \\
 \beta_2\ \mu^2\,& = &\,1.6 \times 10^{-3}\; \; . \nonumber
\end{eqnarray}

In columns 2 and 3 of Table 1, we list the values of the \sub determined
by \S and \G as just described. Ignoring the final column for the moment,
we see that the values for $F^{(0)}_{1,0}$, $\mu^2 F^{(0)}_{1,1}$, 
$F^{(0)}_{2,0}$,
$F^{(2)}_{1,0}$ and $\mu^4 F^{(2)}_{0,2}$ are in close agreement regardless of
which version of $\chi$PT is used. However, each of $F^{(1)}_{1,0}$, 
$\mu^2 F^{(1)}_{1,1}$, $F^{(1)}_{2,0}$, $\mu^4 F^{(1)}_{0,2}$, 
$\mu^2 F^{(2)}_{1,1}$, 
$F^{(2)}_{2,0}$ are predicted to differ by a factor of 2. Consequently, 
one may expect that if we can evaluate these from experiment, data could
distinguish between the two versions of $\chi$PT, at least to 
${\cal O}(p^4)$. It is to the evaluation of these {\it discriminatory} coefficients that we 
now turn.

\begin{table}
\begin{center}
\begin{tabular}{|c|c|c|c|}
\hline
\multicolumn{1}{|c|}{}& \multicolumn{1}{c|}{} & \multicolumn{1}{c|}{} &
\multicolumn{1}{c|}{} \\
Coefficient & S$\chi$PT & G$\chi$PT & Dispersive result (Table 2) \\
& ${\cal O}(p^4)$ & ${\cal O}(p^4)$ & $a_0^0 \, = \, 0.20$ \\ 
& & & \\
\hline
\hline
$F^{(0)}_{1,0}$ & $\;\;\,1.76 \times 10^{-2}$ & $\;\;\,1.72 \times 10^{-2} $ &
$\;\;\,(1.6 \pm 0.2) \times 10^{-2} $\\
\hline
$\mu^2 F^{(0)}_{1,1}$ & $-2.07 \times 10^{-3}$ & $-2.21 \times 10^{-3}$ &
$-(2.1 \pm 0.3) \times 10^{-3}$ \\
\hline
$F^{(0)}_{2,0} $ & $\;\;\,1.23 \times 10^{-3}$ & $\;\;\, 1.42 \times 10^{-3}$ &
$\;\;\,(1.7 \pm 0.3) \times 10^{-3}$ \\
\hline
$F^{(1)}_{1,0} $ & $\;\;\,1.08 \times 10^{-3}$ & $\;\;\, 2.22 \times 10^{-3}$ &
$\;\;\,(2.5 \pm 0.4) \times 10^{-3}$ \\
\hline
$\mu^2 F^{(1)}_{1,1}$ & $-0.51 \times 10^{-3}$ & $-1.18 \times 10^{-3}$ &
$-(1.1 \pm 0.2) \times 10^{-3}$ \\
\hline
$F^{(1)}_{2,0}$ & $\;\;\, 3.32 \times 10^{-4} $& $\;\;\, 7.88 \times 10^{-4}$ &
$\;\;\,(6.8 \pm 1.0)\times 10^{-4}$ \\
\hline
$\mu^4 F^{(1)}_{0,2}$ & $\;\;\, 1.87 \times 10^{-4}$ & $\;\;\,
 4.01 \times 10^{-4}$ &
$\;\;\,(4.8 \pm 0.8)\times 10^{-4}$ \\
\hline
$F^{(2)}_{1,0}$ & $\;\;\, 4.67 \times 10^{-3}$ & $\;\;\, 4.33 \times 10^{-3}$ & 
$\;\;\,(5.1 \pm 0.8) \times 10^{-3}$ \\
\hline
$\mu^2 F^{(2)}_{1,1}$ & $ -1.35 \times 10^{-3}$ & $-1.91 \times 10^{-3}$&
$-(2.1 \pm 0.3) \times 10^{-3}$ \\
\hline
$F^{(2)}_{2,0}$ & $\;\;\, 0.78 \times 10^{-3}$ & $\;\;\, 1.26 \times 10^{-3}$ &
$\;\;\,(1.3 \pm 0.2) \times 10^{-3}$ \\
\hline
$\mu^4 F^{(2)}_{0,2}$ & $\;\;\, 1.23 \times 10^{-3} $ & $\;\;\,
 1.27 \times 10^{-3} $ &
$\;\;\,(1.1 \pm 0.2) \times 10^{-3}$ \\
\hline
\end{tabular}
\end{center}
\caption{\leftskip 1.5cm\rightskip 1.5cm{Comparison of the 
predictions in ${\cal O}(p^4)$ S$\chi$PT and G$\chi$PT for those 
\sub that we shall see,
Sect.~4, can be reliably calculated dispersively, together with the 
results of Table 2.} }
\label{Table1}
\end{table}

\newpage
\baselineskip=7mm
\section{Evaluation of the dispersive integrals~: Test~A}
\hspace*{0.5cm}
The evaluation of the \sub according to Test~A consists
of inputting experimental data for the $\pi \pi$
amplitudes with definite isospin in the $t$--channel
into the dispersive integrals for the \sub, 
Eqs.~(\ref{eq:fit10}--\ref{eq:fit02}), as Chell and Olsson~\cite{mit} did. However
the experimental information in the very low energy region near threshold is still
very poor~\cite{davidmike}. Moreover, as we shall see,
 it is precisely this energy regime that is most important 
 for the evaluation of the \sub. Consequently, we perform
Test~A using a parameterization of the $\pi\pi$ amplitudes
 that reproduces the major features of the experimental data, as a way 
 of restricting the uncertainties.
\par
We calculate the dispersive integrals of 
Eqs.~(\ref{eq:fit10}--\ref{eq:fit02}) by sub-dividing the 
energy range,
$E$, where $s = E^2$, into three regions:
\begin{itemize}
{\leftskip 1.5cm\rightskip 1.5cm{
\item[(I)] $ 2\mu \leq E \leq E_1$, the near threshold region,
\item[(II)] $E_1 < E \leq E_2$, the intermediate energy region,
\item[(III)] $E_2 < E$, the high energy region.}\par}
\end{itemize} 
\noindent $E_1$ is 0.8-0.9~GeV, while
$E_2$ is chosen so that $E_2^2$ is halfway between the $\rho_3(1690)$ and 
$f_4(2050)$
resonance squared masses, in keeping with finite energy sum-rule 
phenomenology,
i.e. $E_2 = 1.85$~GeV.  As we shall see, for almost all the integrals of 
Eqs.~(\ref{eq:fit10}--\ref{eq:fit02}),
region III, where Regge behaviour of the form given in 
Eqs.~(\ref{eq:regge}) applies, gives
a negligible contribution.  We use the Regge residues determined in 
Ref. \cite{mrp}. In region II, the $f_2(1270)$ and $\rho_3(1690)$
contributions are included in the narrow resonance approximation and are also 
for the most part small.
Region I with $E_1 = 0.8$-0.9~GeV generally dominates. In this region,
only $S$ and $P$--waves need be included.  In terms of the phase-shifts, 
$\delta^I_\ell(s)$,
we have
\begin{eqnarray}
{\rm Im} F^{I_t=0} (s,t)\,& = & \,\sqrt{s\over{s-4\mu^2}}\,\Bigg\{
{1\over 3} \sin^2 \delta_0^0(s)\,+\,3 \left(1 + {2t\over{s-4\mu^2}}
\right) \sin^2 \delta^1_1(s)
\,+\,{5\over 3} \sin^2 \delta^2_0(s) \Bigg\}\, , \nonumber \\
& & \nonumber \\
{\rm Im} F^{I_t=1} (s,t)\,& = &\,{4\mu^2\over{2s + t - 4\mu^2}}\,
\sqrt{s\over{s-4\mu^2}}\,\Bigg\{
{1\over 3} \sin^2 \delta_0^0(s)\,+\,{3\over 2} \left(1 + 
{2t\over{s-4\mu^2}}\right)
 \sin^2 \delta^1_1(s) \nonumber \\
& & \qquad \qquad \qquad \qquad \qquad \; \; \; -\,{5\over 6} \sin^2 
\delta^2_0(s) \Bigg\}\, , \\
\label{eq:scha}
& & \nonumber \\
{\rm Im} F^{I_t=2} (s,t)\,& = &\,\sqrt{s\over{s-4\mu^2}}\,\Bigg\{
{1\over 3} \sin^2 \delta_0^0(s)\,-\,{3\over 2} \left(1 + {2t\over{s-4\mu^2}}
\right)
 \sin^2 \delta^1_1(s)
\,+\,{1\over 6} \sin^2 \delta^2_0(s) \Bigg\}\, . \nonumber 
\end{eqnarray}
\baselineskip=7.05mm

\noindent In the low energy region the phase-shifts may usefully be expanded in
powers of momenta by
\begin{equation}
\delta^I_{\ell}(s)\,=\,\left({s-4\mu^2\over {4\mu^2}}\right)^{\ell+1/2}\,
\left[ \,a^I_{\ell}\,
+\,b^I_{\ell} \left({s-4\mu^2\over {4\mu^2}}\right)+\,...\,\right]\; ,
\label{eq:phases}
\end{equation}
\noindent where $a^I_{\ell}$ are the
 scattering lengths and $b^I_{\ell}$ the effective ranges.
This near threshold expansion is naturally
embodied in the following flexibly convenient representation of the 
phase-shifts \cite{schenkk} in terms of the $K$--matrix
\begin{eqnarray}
K^I_{\ell} \, \equiv \, \sqrt{\Frac{s}{s \, - \, 4 \mu^2}} \,  
\tan \delta^I_{\ell} (s)\,& = &\,
\left({s-4\mu^2\over {4\mu^2}}\right)^{\ell}\,\Bigg\{a^I_{\ell}\,+\,
\tilde{b}_{\ell}^I\,\left({s-4\mu^2\over {4\mu^2}}\right)
\Bigg\}\,{{4\mu^2-s^I_{\ell}}
\over{s-s^I_{\ell}}}\, , \nonumber \\
& & \label{eq:schenk} \\
\tilde{b}_{\ell}^I\,& = & \,b_{\ell}^I\,-\,
a^I_{\ell}\left(4\mu^2\over{s^I_l-4\mu^2}\right)\,
+\,(a^I_{\ell})^3\,\delta_{\ell 0}\, ,\nonumber 
\end{eqnarray}
\noindent where, as already mentioned, the $a^I_{\ell}$,  $b^I_{\ell}$ and $s^I_{\ell}$ are 
fixed to give a
parameterization consistent with experiment and with $\chi$PT
 in its appropriate version. Other parameterizations have been tried
 and these alter our numbers little --- this is a consequence of
  the integrals being dominated by the near 
 threshold absorptive parts.  To illustrate this, we show in Fig.~1, the
 integrands for
$F^{(I_t)}_{k,m}$
\noindent as functions of energy $E= \sqrt s$ for two different values of the
$I=0$ $S$--wave scattering length $a^0_0$. One sees that the low 
energy region largely determines their dispersive evaluation, except for
$F^{(1,2)}_{0,1}$.

In Table 2, we present the contributions to the dispersive integrals
in regions I, II and III. 
 In region I the $S$--wave parameters have been fixed to
those determined by Schenk \cite{schenkk}, which represent the 
well-known experimental results
reviewed in \cite{davidmike}
 and match one loop S$\chi$PT near threshold, i.e.
we take $\mu=139.6$~MeV, $a_0^0 = 0.20$, $b^0_0 = 0.24$, $s_0^0 = (0.865\, 
{\rm GeV})^2$,
$a^2_0 = -0.042$, $b^2_0 = -0.075$, $s^2_0 = -(0.685\, {\rm GeV})^2$.
For the $P$--wave we take $a_1^1 = 0.037$, $b^1_1 =0.005$ and 
$s^1_1 = (0.77\, {\rm GeV})^2$, the squared $\rho-$mass. The uncertainties
are typically 10\% in region II, 25\% in region III and,
even keeping $a^0_0$ fixed at its ${\cal O}(p^4)$
$\chi$PT value, $\sim 15\%$ in region I.
With these uncertainties, the dispersive results of Table 2 are added
as the last column in Table 1. One now sees that the coefficients, 
for which \S and \G predicted a common value, agree well with their 
dispersive evaluation from experiment. In contrast, the so called
{\em discriminatory} coefficients, that generally differ by a factor
of 2 at ${\cal O}(p^4)$, are far closer to the predictions of \G 
(with $\alpha=3.1$, $\beta=0.93$) than with \S. This is at first
sight rather surprising since the input absorptive parts have been
explicitly designed (by Schenk \cite{schenkk}) to match ${\cal O}(p^4)$ \S.
In contrast, the coefficients in \G to ${\cal O}(p^4)$ are in very good
agreement with the same dispersive evaluation. If this were the whole
story then this would indicate that low orders in \G more rapidly
embody key resonance contributions.
\par

\begin{table}
\begin{center}
\begin{tabular}{|c|c|c|c|c|}
\hline
\multicolumn{1}{|c|}{} & \multicolumn{1}{|c|}{} & \multicolumn{1}{|c|}{} &
\multicolumn{1}{|c|}{} & \multicolumn{1}{|c|}{} \\ 
\multicolumn{1}{|c|}{Coefficient} &
\multicolumn{1}{|c|}{Region I} &
\multicolumn{1}{c|}{Region II} &
\multicolumn{1}{c|}{Region III} & 
\multicolumn{1}{|c|}{Total} \\
& & & & \\
\hline
\hline
$F^{(0)}_{1,0}$ &$\, \; \; 1.56 \times 10^{-2}$ &$ \;\, 3.46 \times 10^{-4}$ &
$2.54 \times 10^{-4} $ & $\,\;\;1.6 \times 10^{-2}$\\
\hline
$\mu^2 F^{(0)}_{1,1}$ & $-2.19 \times 10^{-3}$ &$ \;\, 2.09 \times 10^{-5}$ &
$1.63 \times 10^{-5}$ & $-2.1 \times 10^{-3}$\\
\hline
$F^{(0)}_{2,0}$ & $\;\, \; 1.67 \times 10^{-3}$ & $\;\, 1.92 \times 10^{-7}$ & 
$1.10 \times 10^{-8}$ & $\;\;\,1.7 \times 10^{-3}$\\
\hline
$F^{(1)}_{1,0}$ & $\;\, \;2.52 \times 10^{-3}$ & $\;\, 6.96 \times 10^{-6}$ &
$5.30 \times 10^{-7} $ & $\;\;\,2.5 \times 10^{-3}$\\
\hline
$\mu^2 F^{(1)}_{0,1}$ & $-1.46 \times 10^{-3}$ & $\;\, 9.57 \times 10^{-4}$ & 
$2.00 \times 10^{-3} $ &\, unreliable\\
\hline
$\mu^2 F^{(1)}_{1,1}$ & $-1.08 \times 10^{-3}$ & $\;\, 4.05 \times 10^{-7}$ & 
$3.65 \times 10^{-8} $ & $-1.1 \times 10^{-3}$\\
\hline
$F^{(1)}_{2,0}$ & $\; \; \,6.79 \times 10^{-4}$ & $\;\, 4.34 \times 10^{-9}$ & 
$\,3.81 \times 10^{-11}$ & $\;\;\,6.8 \times 10^{-4}$\\
\hline
$\mu^4 F^{(1)}_{0,2}$ &$\; \; \, 3.86 \times 10^{-4}$ & $\;\, 8.78 \times 10^{-6}$ &
$9.12 \times 10^{-5}$ & $\;\;\,4.8 \times 10^{-4}$\\
\hline
$F^{(2)}_{1,0}$ & $\; \; \,4.88 \times 10^{-3}$ & $\;\, 2.29 \times 10^{-4}$ & 
${\cal O}(10^{-5})$ & $\;\;\,5.1 \times 10^{-3}$\\
\hline
$\mu^2 F^{(2)}_{0,1}$ & $-3.98 \times 10^{-2}$ & $\;\, 1.39 \times 10^{-2}$ & 
${\cal O}(10^{-2})$ &\, unreliable\\
\hline
$\mu^2 F^{(2)}_{1,1}$ & $-2.12 \times 10^{-3}$ & $\;\, 1.25 \times 10^{-5}$ &
${\cal O}(10^{-6})$ & $-2.1 \times 10^{-3}$\\
\hline
$F^{(2)}_{2,0}$ & $\; \; \,1.25 \times 10^{-3}$ & $\;\, 1.70 \times 10^{-7}$ &
${\cal O}(10^{-8})$ & $\;\;\,1.3 \times 10^{-3}$\\
\hline
$\mu^4 F^{(2)}_{0,2}$ & $\;\;\,1.13 \times 10^{-3}$ & $-1.00 \times 10^{-5}$ &
${\cal O}(10^{-5})$ &  $\;\;\,1.1 \times 10^{-3}$ \\
\hline
\end{tabular}
\end{center}
\caption{\leftskip 1.5cm\rightskip 1.5cm
{Contribution to the subthreshold coefficients Eqs.~(\protect 
\ref{eq:fit10}-\protect \ref{eq:fit02})
from the three different energy regions as explained in the text.
Typical uncertainties are 15\%, 10\%, 25\%, respectively, in these
three contributions.}}
\label{Table2}
\end{table} 
However, let us return to the evaluation of the {\em discriminatory} 
coefficients from experimental information. In Fig. 1 we see that all
the {\em discriminatory} coefficients are entirely dominated by the 
very near threshold region below $450$~MeV or so and though the values
given in column 4th of Table 1 have $15\%$ errors, this is assuming
a particular value of the $I=0$ $S$--wave scattering length, 
$a_0^0 = 0.20$ in Eq.~(\ref{eq:schenk}). If we fold in the real 
uncertainties from the Geneva--Saclay $K_{e4}$ results~\cite{GSaclay} on the near
threshold phase--shifts, then one would readily see that for these
coefficients the present experimental uncertainties are more than
$100 \%$~\cite{davidmike} (compare the two curves in Fig.~1) encompassing both \S and what we call \G in Table 1.
 Thus
experiment cannot presently distinguish between these differing 
versions of $\chi$PT and so Test~A is inconclusive. Nonetheless, the
Chell and Olsson test in form B does tell us that the ${\cal O}(p^6)$
corrections in \S must be large, as we next discuss. 
\section{When is the Chell--Olsson test an identity (Test~B)~?}
\hspace*{0.4cm}
\baselineskip=7.mm
For the {\em discriminatory} coefficients, the dispersive integrals
are controlled by the near threshold region where we would expect 
$\chi$PT should itself be applicable. Then these relations should be
an identity, since the amplitudes of $\chi$PT satisfy the crossing
and analyticity properties that Eqs.~(\ref{eq:f12dr},\ref{eq:f0dr})
embody. Thus if we evaluate the \sub directly from $\chi$PT in either
form at ${\cal O}(p^4)$, for example, or alternatively input the
${\cal O}(p^4)$ imaginary parts into the dispersive integrals for 
$F^{(I_t)}_{k,m}$, Eqs.~(\ref{eq:fit10}--\ref{eq:fit02}), the results
should be the same. However, comparing columns 2 and 4 of Table 1, 
we see that inputting phases with the ${\cal O}(p^4)$ $S$--wave scattering
length of $a_0^0 = 0.20$ does not reproduce \S. This is because
$\chi$PT only satisfies unitarity perturbatively. Consistency with the
${\cal O}(p^4)$ \sub requires the ${\cal O}(p^4)$ imaginary parts
be input into the dispersive integrals. This ${\cal O}(p^4)$ absorptive
part is wholly given by the ${\cal O}(p^2)$ real part, since
\begin{equation}
{\rm Im} \, f_{\ell}^I (s, {\cal O}(p^4)) \; = \; 
\sqrt{1- \Frac{4 \mu^2}{s}} \, \left[ \, {\rm Re} \, f_{\ell}^I 
(s,{\cal O}(p^2)) \, \right]^2 \; \; \; \; . 
\label{eq:uniper}
\end{equation}
Thus the Chell--Olsson test should become an identity, if we input 
phases at the appropriate order. Working to ${\cal O}(p^4)$ for the
coefficients, we must input phases at ${\cal O}(p^2)$ for which
$a_0^0 = 0.16$ in \S or $a_0^0 = 0.23$ in our version of \G with
$\alpha=3.1$. With these values we obtain the results in Table 3. 
\par
We now see complete agreement between the dispersive results and the 
explicit evaluation (except for $F^{(2)}_{1,0}$ which appears to 
be due to a poorer convergence in the dispersive 
evaluation). The fact that the discriminatory
coefficients in \S and \G differ by a factor 2 at ${\cal O}(p^4)$
just reflects the 
fact that the imaginary parts of the near threshold amplitudes
are very nearly proportional to $(a_0^0)^2$ and $(0.23/0.16)^2 \, 
\approx \, 2.1$. Moreover, the large change in the {\em discriminatory}
coefficients in \S between their values in column 2 of Table 3 and
column 2 of Table 1 means that the ${\cal O}(p^6)$ corrections
must be large --- since inputting ${\cal O}(p^4)$ phases in Table 1 
generates ${\cal O}(p^6)$ coefficients. The recent two loop calculation
of the $\pi \pi$ amplitude by Bijnens et al.~\cite{bce96} bears this out, 
as Moussallam has checked \footnote{B. Moussallam, private communication.}.
In Table 4 we give examples of the change at ${\cal O}(p^6)$, for 
$F^{(1)}_{1,0}$ and $F^{(1)}_{1,1}$. The self--consistency is now clear
after these $100 \%$ corrections.
\par

\begin{table}
\begin{center}
\begin{tabular}{||c||c|c||c|c||}
\hline
\multicolumn{1}{||c||}{}& \multicolumn{1}{c|}{} & \multicolumn{1}{c||}{} &
\multicolumn{1}{c|}{} & \multicolumn{1}{c||}{}\\
Coefficient & S$\chi$PT & Dispersive results & G$\chi$PT & Dispersive results \\
&${\cal O}(p^4)$ &$a^0_0=0.16$ &${\cal O}(p^4)$  & $a^0_0=0.23$ \\
& & & & \\
\hline
\hline
$F^{(0)}_{1,0}$ & $\;\;\,1.76 \times 10^{-2}$ & 
$\;\;\,(1.4 \pm 0.2) \times 10^{-2} $ &
$\;\;\,1.72 \times 10^{-2} $  & $\;\;\,(1.4 \pm 0.2) \times 10^{-2} $\\
\hline
$\mu^2 F^{(0)}_{1,1}$ & $-2.07 \times 10^{-3}$ &
$-(1.5 \pm 0.3) \times 10^{-3}$ & 
$-2.21 \times 10^{-3}$  & $-(1.8 \pm 0.4) \times 10^{-3} $ \\
\hline
$F^{(0)}_{2,0} $ & $\;\;\,1.23 \times 10^{-3}$ &
$\;\;\,(1.3 \pm 0.3) \times 10^{-3}$ &
 $\;\;\, 1.42 \times 10^{-3}$  & $\;\;\,(1.5 \pm 0.3) \times 10^{-3} $ \\
\hline
$F^{(1)}_{1,0} $ & $\;\;\,1.08 \times 10^{-3}$ &
 $\;\;\,(1.5 \pm 0.3) \times 10^{-3}$ &
$\;\;\, 2.22 \times 10^{-3}$ &
 $\;\;\,(2.6 \pm 0.4) \times 10^{-3} $\\
\hline
$\mu^2 F^{(1)}_{1,1}$ & $-0.51 \times 10^{-3}$ &
$-(0.6 \pm 0.1) \times 10^{-3}$ &
 $-1.18 \times 10^{-3}$ & $-(1.2 \pm 0.3) \times 10^{-3} $ \\
\hline
$F^{(1)}_{2,0}$ & $\;\;\, 3.32 \times 10^{-4} $&
$\;\;\,(3.7 \pm 0.6)\times 10^{-4}$ &
 $\;\;\, 7.88 \times 10^{-4}$ 
& $\;\;\,(8.2 \pm 0.5) \times 10^{-4} $ \\
\hline
$\mu^4 F^{(1)}_{0,2}$ & $\;\;\, 1.87 \times 10^{-4}$ &
$\;\;\,(2.0 \pm 0.5)\times 10^{-4}$ &
 $\;\;\, 4.01 \times 10^{-4}$ & $\;\;\,(4.0 \pm 0.4) \times 10^{-4} $\\
\hline
$F^{(2)}_{1,0}$ & $\;\;\, 4.67 \times 10^{-3}$ &
$\;\;\,(2.6 \pm 0.6) \times 10^{-3}$ &
 $\;\;\, 4.33 \times 10^{-3}$ &  $\;\;\,(3.9 \pm 0.6) \times 10^{-3} $\\
\hline
$\mu^2 F^{(2)}_{1,1}$ & $ -1.35 \times 10^{-3}$ &
$-(1.4 \pm 0.2) \times 10^{-3}$ &
 $-1.91 \times 10^{-3}$& $-(2.0 \pm 0.4) \times 10^{-3} $ \\
\hline
$F^{(2)}_{2,0}$ & $\;\;\, 0.78 \times 10^{-3}$ &
$\;\;\,(0.8 \pm 0.1) \times 10^{-3}$ &
 $\;\;\, 1.26 \times 10^{-3}$ & $\;\;\,(1.3 \pm 0.2) \times 10^{-3} $ \\
\hline
$\mu^4 F^{(2)}_{0,2}$ & $\;\;\, 1.23 \times 10^{-3} $ &
$\;\;\,(1.0 \pm 0.2) \times 10^{-3}$ &
 $\;\;\, 1.27 \times 10^{-3} $ & $\;\;\,(1.1 \pm 0.2) \times 10^{-3} $ \\
\hline
\end{tabular}
\end{center}
\caption{\leftskip 1.5cm\rightskip 1.5cm{Comparison of the 
results for those subthreshold coefficients
that can be reliably calculated dispersively, with $a_0^0 = 0.16$ and 0.23,
 with their
predictions in S$\chi$PT and G$\chi$PT at ${\cal O}(p^4)$.}}
\label{Table3}
\end{table}
 
The fact that the ${\cal O}(p^6)$ $I=0$ $S$--wave scattering length differs by 
$8 \%$ from its ${\cal O}(p^4)$ value in 
\S \cite{bce96}, allows us to estimate that the {\em discriminatory} 
coefficients
will have just a $17 \%$ correction at ${\cal O}(p^8)$ and so the 
Standard perturbative expansion is improving.

\begin{table}
\begin{center}
\begin{tabular}{|c|c|c|c|}
\hline
\multicolumn{1}{|c|}{Coefficient} & 
\multicolumn{1}{c|}{\S $\, {\cal O}(p^4)$} & 
\multicolumn{1}{c|}{\S $\, {\cal O}(p^6)$} &
\multicolumn{1}{c|}{Dispersive ($a_0^0 = 0.20$)} \\
\hline
\hline
$ F^{(1)}_{1,0} \times 10^3$ & $\; \; \, 1.08 $ & $\, \; \; 2.51 $ & 
$\, \; \; (2.5 \pm 0.4)$ \\
\hline
$\mu^2 F^{(1)}_{1,1} \times 10^3$ & $-0.51$ & $ -0.95$ & $-(1.1 \pm 0.2)$\\
\hline
\end{tabular}
\end{center}
\caption{
\leftskip 1.5cm\rightskip 1.cm{Comparison between the \S evaluation at 
${\cal O}(p^4)$, ${\cal O}(p^6)$ and the dispersive calculation for 
$a_0^0=0.20$ of two of the {\em discriminatory} coefficients. The 
${\cal O}(p^6)$ \S computation was made by Moussallam~$^4$.}}
\label{Table4}
\end{table}

\newpage
\section{Conclusions}
\hspace*{0.5cm}
\baselineskip=7.5mm
The Chell--Olsson test is indeed stringent. However, using presently
available experimental information, it is not able to distinguish
between \S and \G. This reflects the large uncertainties in the
near threshold $S$--wave phases that hopefully measurements of 
$K_{e4}$ decays with higher statistics and smaller systematic
uncertainties at DA$\Phi$NE will improve.
\par
The coefficients in the subthreshold expansion that
have most potential to distinguish both forms of $\chi$PT, the ones
we have called {\em discriminatory}, all have no polynomial 
${\cal O}(p^4)$ corrections in terms of the $\overline{\ell}_i$'s of
\S. Curiously enough we have shown that these same coefficients have
$\sim 100 \%$ corrections at ${\cal O}(p^6)$ and the recent explicit
calculation of Bijnens et al.~\cite{bce96} of the two loop $\pi \pi$
amplitude shows that this is in fact so. Indeed, these same calculations
and our study allow an estimate of $\sim 17 \%$  to be made 
for the ${\cal O}(p^8)$ corrections to these \sub.
Thus the Chell--Olsson test becomes an identity when applied to the
amplitudes of $\chi$PT. This is because they have the crossing and
analytic properties of the full amplitude provided we recognize
that unitarity is only fulfilled perturbatively --- order by order.

\vspace*{1.5cm} 
\noindent {\Large{Acknowledgements}}
\vspace*{0.2cm}

This collaboration was made possible by the 
 support of the EC Human Capital and Mobility Programme which funds the
{\it EURODA$\Phi$NE} network under grant CHRX-CT920026.
JP wishes to thank the Particle Physics Department of the Rutherford--Appleton
Laboratory, where this work was started, for their kind hospitality.
We are grateful to Gilberto Colangelo, J\"urg Gasser, Marc Knecht and 
Jan Stern for discussions and particularly to Bachir Moussallam for his
evaluation of the coefficients in two loops \S. We are also grateful
to Hans Bijnens and Ulf-G. Meissner for the opportunity to present
this work at the ECT*{\it Workshop on the Standard Model at Low Energies} at 
Trento in May 1996. J.P. is also partially supported by DGICYT under 
grant PB94--0080.
 
\vspace*{1.cm}
\noindent{\Large{Figure Captions}} 
\vspace*{0.5mm}

\noindent{\bf Figure 1} : Integrands of the dispersion relations for 
$F^{(I_t)}_{k,m}$,
as in
Eqs.~(\ref{eq:fit10}-\ref{eq:fit02}), in energy region I as function of 
$\sqrt s$ evaluated using Eqs.~(49,47)
with
$a^0_0\,=\,0.20$ (solid lines) and $a^0_0\,=\,0.27$ (dashed lines). 
The ordinate is given in  arbitrary dimensionless units.

\newpage

\pagestyle{empty}
\thispagestyle{empty}

\begin{figure}
\begin{center}
\leavevmode
\hbox{%
\epsfxsize=16cm
\epsffile{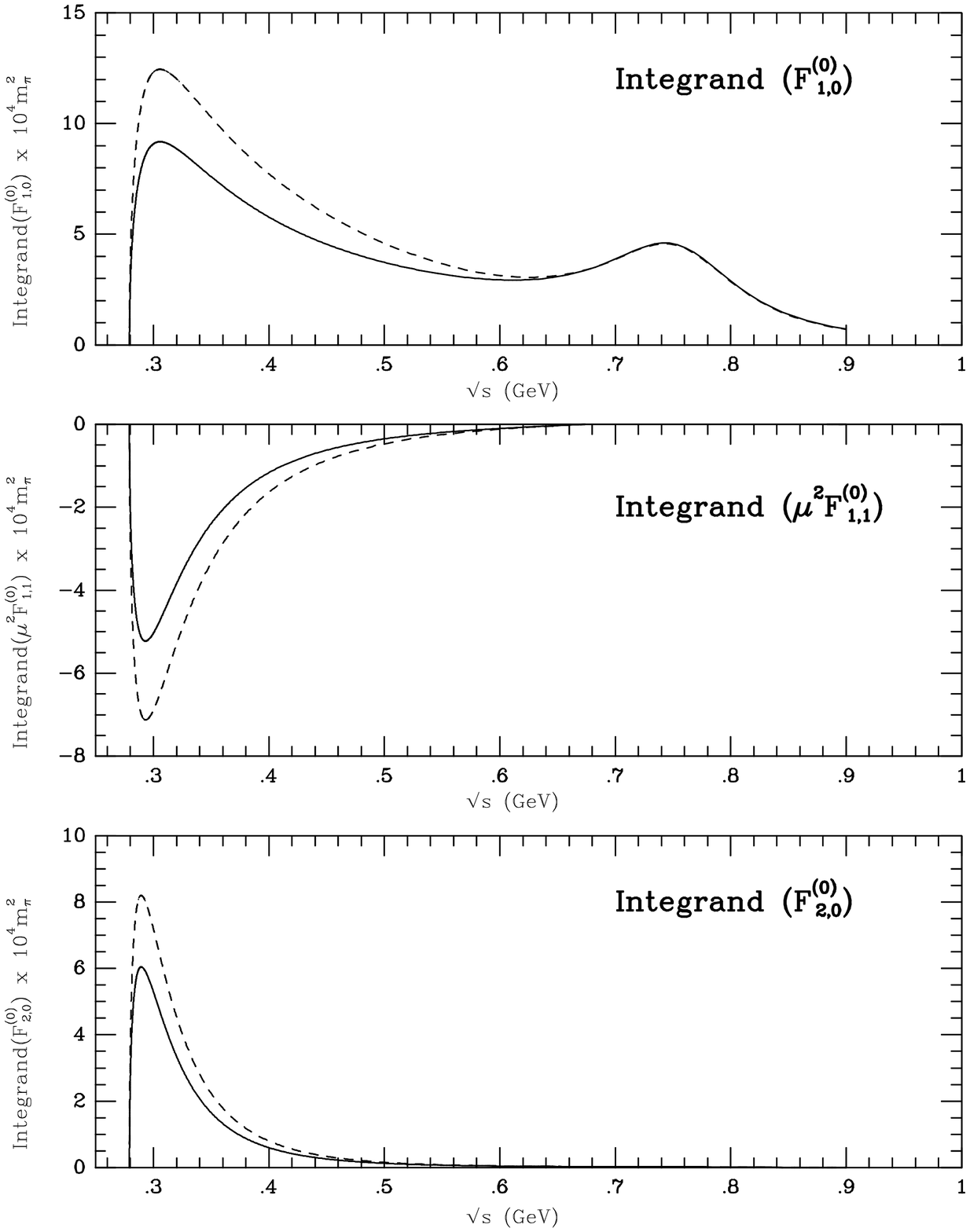}}
\end{center}
\end{figure}

\begin{figure}
\begin{center}
\leavevmode
\hbox{%
\epsfxsize=16cm
\epsffile{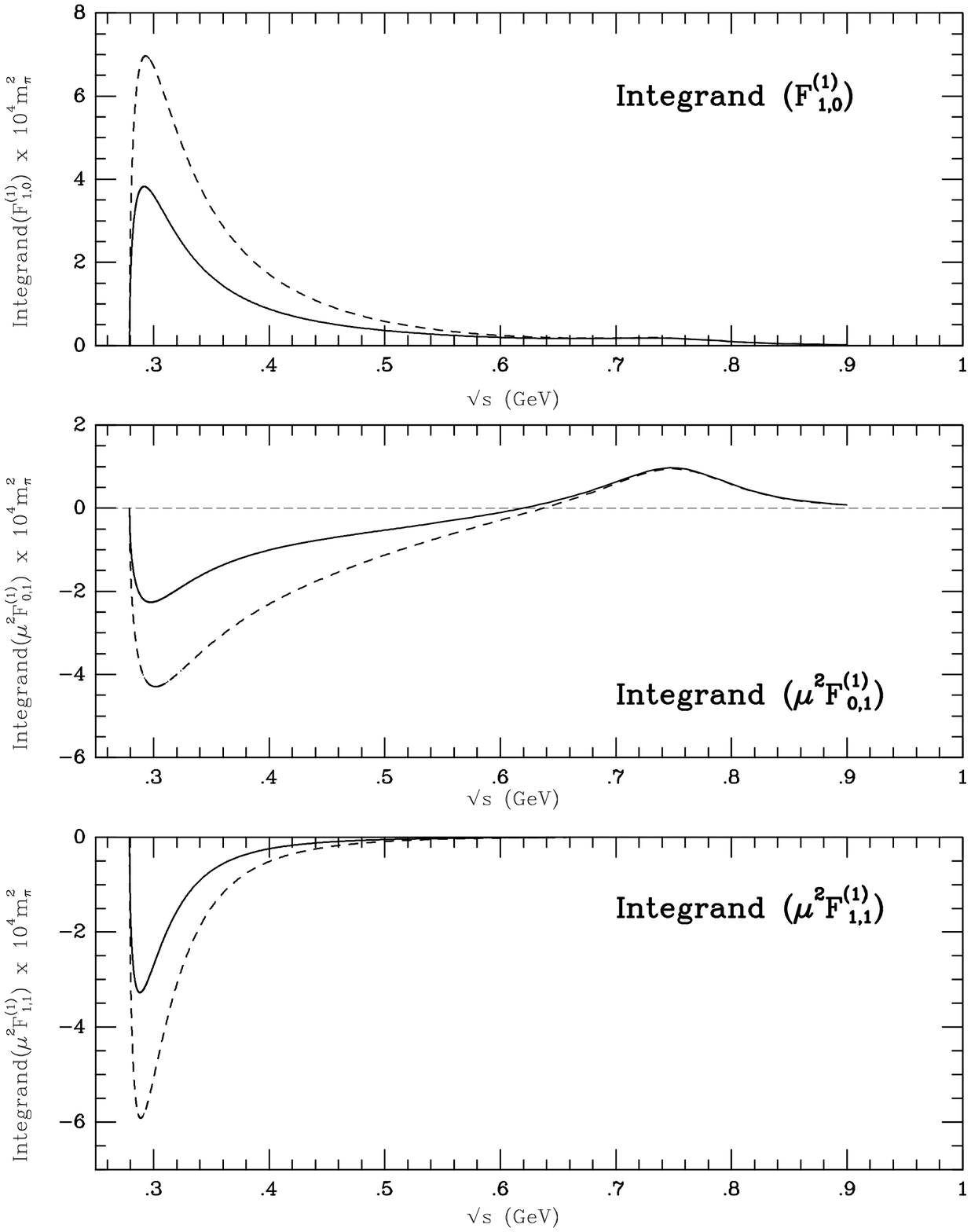}}
\end{center}
\end{figure}

\begin{figure}
\begin{center}
\leavevmode
\hbox{%
\epsfxsize=16cm
\epsffile{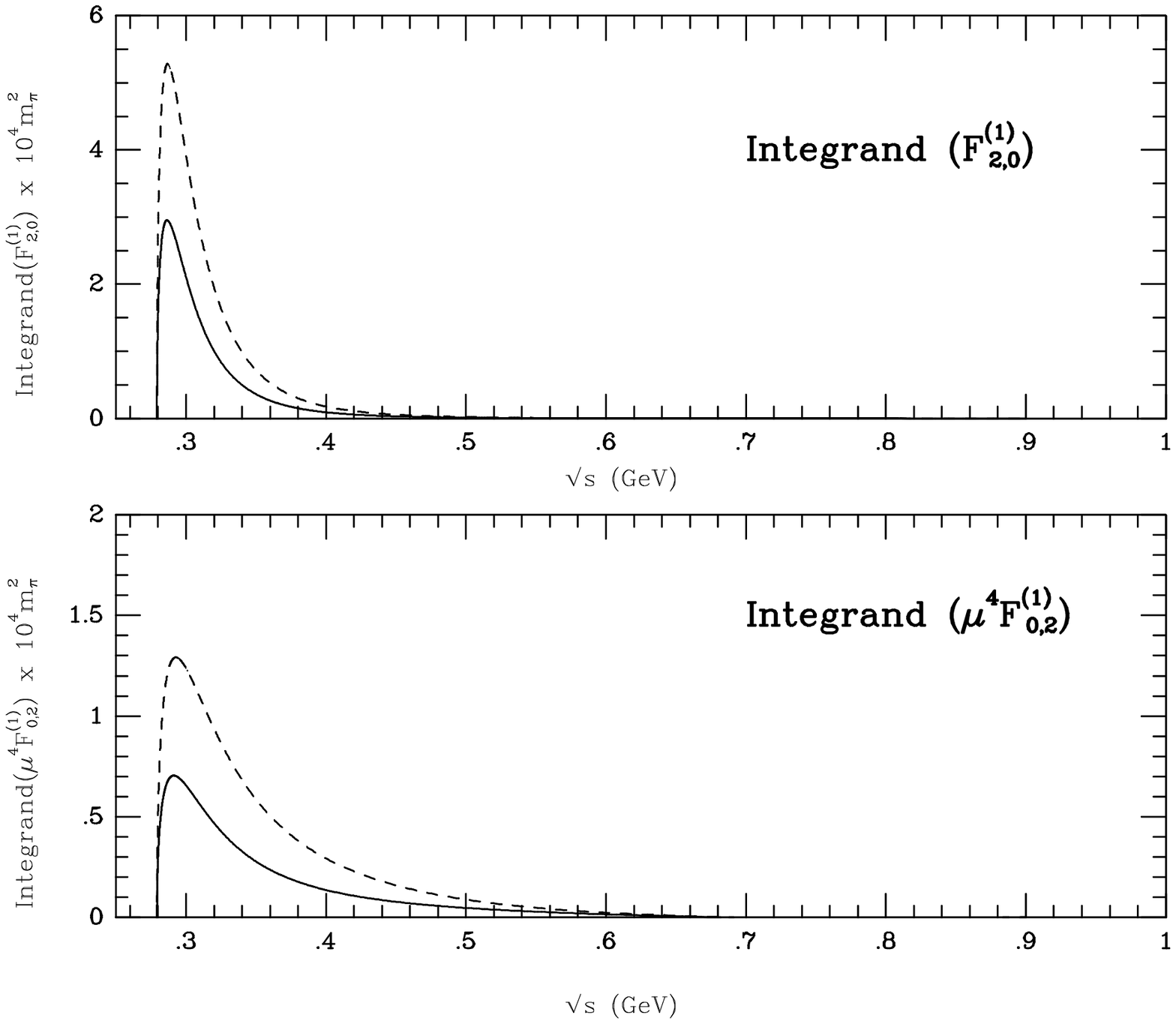}}
\end{center}
\end{figure}

\begin{figure}
\begin{center}
\leavevmode
\hbox{%
\epsfxsize=16cm
\epsffile{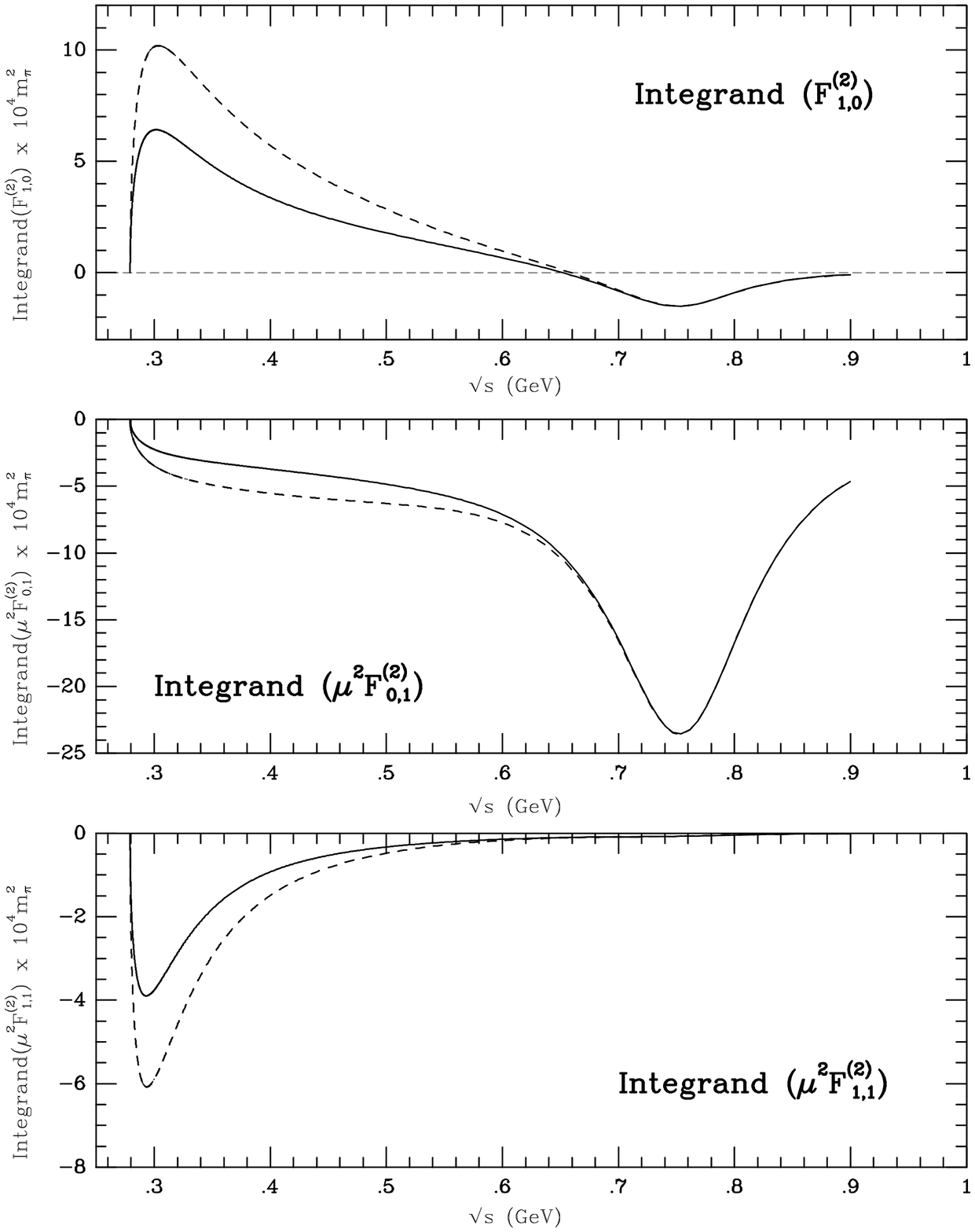}}
\end{center}
\end{figure}

\begin{figure}
\begin{center}
\leavevmode
\hbox{%
\epsfxsize=16cm
\epsffile{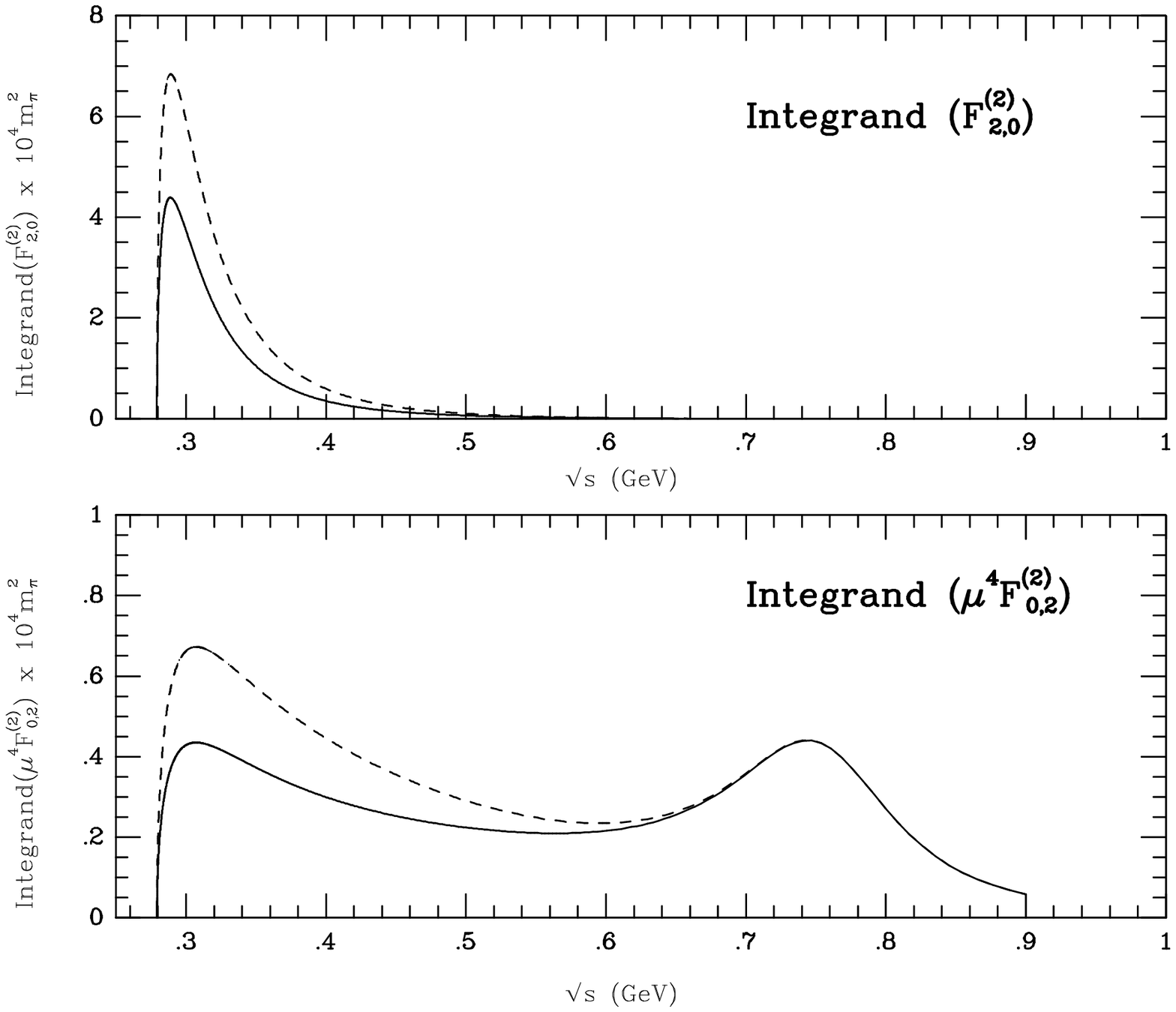}}
\end{center}
\end{figure}

\end{document}